\documentclass[a4paper,12pt]{article} 
\usepackage{color}
\usepackage{epsfig}
\usepackage{amsfonts}
\usepackage{amsmath}
\usepackage{amssymb}
\usepackage{rotating}
\usepackage{multirow}
\usepackage{mathtools}
\usepackage{chngpage}
\usepackage{setspace}
\usepackage{array}
\usepackage{url}
\usepackage{booktabs}
\usepackage[square, authoryear]{natbib}
\usepackage{array}
\usepackage{caption}
\DeclareMathOperator{\se}{se}
\makeatletter
\renewcommand\@biblabel[1]{}
\makeatother
\renewcommand{\thetable}{\Roman{table}}

\begin{document}
\title{Mendelian randomization with fine-mapped genetic data: choosing from large numbers of correlated instrumental variables}
\author{}
\author{Stephen Burgess \textsuperscript{1,2} \thanks{Corresponding author: Dr Stephen Burgess. Address: MRC Biostatistics Unit, Cambridge Institute of Public Health, Robinson Way, Cambridge, CB2 0SR, UK. Telephone: +44 1223 748651. Fax: none. Email: sb452@medschl.cam.ac.uk.} \and Verena Zuber \textsuperscript{3} \and Elsa Valdes-Marquez \textsuperscript{4} \and Benjamin B Sun \textsuperscript{2} \and Jemma C Hopewell \textsuperscript{4}  \\ \\
\textsuperscript{1} MRC Biostatistics Unit, Cambridge, UK \\
\textsuperscript{2} Department of Public Health and Primary Care, \\ University of Cambridge, UK \\
\textsuperscript{3} European Bioinformatics Institute, Hinxton, nr Duxford, UK \\
\textsuperscript{4} Clinical Trial Services Unit, Nuffield Department of Population Health, \\ BHF Centre for Research Excellence, University of Oxford, UK}
\maketitle 

\noindent \noindent \textbf{Running title:} Mendelian randomization with fine-mapped data [45 characters].

\clearpage

\setstretch{1.4}
\subsection*{Abstract}
Mendelian randomization uses genetic variants to make causal inferences about the effect of a risk factor on an outcome. With fine-mapped genetic data, there may be hundreds of genetic variants in a single gene region any of which could be used to assess this causal relationship. However, using too many genetic variants in the analysis can lead to spurious estimates and inflated Type 1 error rates. But if only a few genetic variants are used, then the majority of the data is ignored and estimates are highly sensitive to the particular choice of variants. We propose an approach based on summarized data only (genetic association and correlation estimates) that uses principal components analysis to form instruments. This approach has desirable theoretical properties: it takes the totality of data into account and does not suffer from numerical instabilities. It also has good properties in simulation studies: it is not particularly sensitive to varying the genetic variants included in the analysis or the genetic correlation matrix, and it does not have greatly inflated Type 1 error rates. Overall, the method gives estimates that are not so precise as those from variable selection approaches (such as using a conditional analysis or pruning approach to select variants), but are more robust to seemingly arbitrary choices in the variable selection step. Methods are illustrated by an example using genetic associations with testosterone for 320 genetic variants to assess the effect of sex hormone-related pathways on coronary artery disease risk, in which variable selection approaches give inconsistent inferences.

\vspace{2mm}

\noindent \noindent \textbf{Keywords:} Mendelian randomization, allele score, correlated variants, summarized data, conditional analysis.

\clearpage

\setstretch{1.4}
\section*{Background}
In a Mendelian randomization investigation, genetic variants that are instrumental variables for a given risk factor are used to assess the causal effect of the risk factor on an outcome \citep{daveysmith2003, burgess2015book}. An association between such a genetic variant and the outcome is indicative of a causal effect of the risk factor on the outcome \citep{didelez2007, lawlor2007}. When there are multiple uncorrelated genetic variants that are instrumental variables for the same risk factor, power to detect a causal effect is maximized by including all such genetic variants in a single analysis \citep{pierce2010}. However, when genetic variants are correlated, it is not clear how to choose which variants to include in the analysis to obtain the most efficient estimate possible without the analysis suffering from numerical instabilities when there are large numbers of highly-correlated candidate variants (such as with fine-mapped genetic data).

\subsection*{Theoretical viewpoint}
If individual-level data are available on the genetic variants (potentially correlated), risk factor, and outcome for the same participants, then the two-stage least squares (2SLS) method provides the most efficient estimate of the causal effect (amongst all instrumental variable estimators using linear combinations of the instruments and under conditional homoscedasticity -- the error term in the model relating the outcome to the risk factor has constant variance conditional on the instruments) \citep{wooldridge2009ch15}.

The first stage of the 2SLS method regresses the risk factor on all the genetic variants. As the sample size increases, the coefficient of any variant that does not explain independent variation in the risk factor will tend to zero, and so its contribution to the analysis decreases to zero. This implies that an optimally-efficient Mendelian randomization analysis should include all genetic variants associated with the risk factor in a conditional analysis. The inclusion of additional variants not independently associated with the risk factor will not have a negative impact on the analysis asymptotically (as their coefficient for contribution to the analysis will tend to zero), but will not add to the precision of the causal estimate either. As an aside, fitted values from the first-stage of the 2SLS method are equivalent (up to an additive constant) to values of an allele score (also called a genetic risk score). This implies that the optimal weights in an allele score with correlated variants are the conditional (multivariable) associations of the variants with the risk factor.

\subsection*{Estimating a causal effect using summarized data}
The 2SLS estimate can also be obtained using summarized data on genetic associations with the risk factor and with the outcome from univariable regression analyses of the risk factor or outcome on each genetic variant in turn. This is important as such summarized data from large consortia are often publicly available, enabling Mendelian randomization investigations to be performed on large sample sizes without the need for costly and time-consuming data-sharing arrangements \citep{burgess2014twosample}. This estimate can also be calculated in a two-sample setting, in which genetic associations with the risk factor and with the outcome are estimated in different samples \citep{inoue2010}.

If the genetic association with the risk factor for genetic variant $j$ is $\hat{\beta}_{Xj}$ with standard error $\se(\hat{\beta}_{Xj})$, and with the outcome is $\hat{\beta}_{Yj}$ with standard error $\se(\hat{\beta}_{Yj})$, and assuming that genetic variants are uncorrelated, then the causal estimate is \citep{johnson2013vig}:
\begin{equation}
\mbox{Inverse-variance weighted estimate (uncorrelated variants) = } \frac{\sum_j \hat{\beta}_{Yj} \hat{\beta}_{Xj} \se(\hat{\beta}_{Yj})^{-2}}{\sum_j \hat{\beta}_{Xj}\hphantom{}^2 \se(\hat{\beta}_{Yj})^{-2}}. \label{eq:ivw}
\end{equation}
This is referred to as the inverse-variance weighted (IVW) estimate \citep{burgess2013genepi}. It is the weighted mean of the 2SLS estimates using each genetic variant individually ($\frac{\hat{\beta}_{Yj}}{\hat{\beta}_{Xj}}$) with the inverse-variance weights $\left[\frac{\se(\hat{\beta}_{Yj})}{\hat{\beta}_{Xj}}\right]^{-2}$. The variant-specific estimates are combined using the standard formula for a fixed-effect meta-analysis \citep{borenstein2009}. This same estimate can be obtained by weighted regression of the genetic associations with the outcome $\hat{\beta}_{Yj}$ on the genetic associations with the risk factor $\hat{\beta}_{Xj}$ using weights $\se(\hat{\beta}_{Yj})^{-2}$ and with the intercept term set to zero. The IVW estimate is equivalent to the 2SLS estimate when the genetic variants are uncorrelated \citep{burgess2014pleioajeappendix}. This formula does not take into account uncertainty in the genetic associations with the risk factor; however, these associations are typically more precisely estimated than those with the outcome, and ignoring this uncertainty does not lead to inflated Type 1 error rates for the IVW estimate in realistic scenarios \citep{burgess2013genepi}.

When genetic variants are correlated, the IVW method can be extended to account for the correlations between genetic variants \citep{burgess2015scoretj}. This can be motivated by considering generalized weighted linear regression of the genetic associations with the outcome on the genetic associations with the risk factor using the weighting matrix $\Omega$, where $\Omega_{j_1 j_2} = \se(\hat{\beta}_{Yj1}) \se(\hat{\beta}_{Yj2}) \rho_{j_1 j_2}$, and $\rho_{j_1 j_2}$ is the correlation between genetic variants $j_1$ and $j_2$. The causal estimate is:
\begin{equation}
\mbox{Inverse-variance weighted estimate (correlated variants) = } (\hat{\boldsymbol\beta}_{X}^T \Omega^{-1} \hat{\boldsymbol\beta}_{X})^{-1} \hat{\boldsymbol\beta}_{X}^T \Omega^{-1} \hat{\boldsymbol\beta}_{Y} \label{eq:ivwgen}
\end{equation}
where $\hat{\boldsymbol\beta}_{X}, \hat{\boldsymbol\beta}_{Y}$ are vectors of the genetic associations, and $^T$ is a vector transpose. Again, this estimate is equivalent to the 2SLS estimate that is obtained using individual-level data (see Appendix for proof). It therefore inherits the efficiency property of the 2SLS estimate as the optimally-efficient causal estimate based on all the genetic variants.

\subsection*{Scope of paper}
In this paper, we illustrate and provide guidance on choosing variants to include in a Mendelian randomization with fine-mapped genetic data. We first provide a motivating example analysis based on summarized genetic associations for hundreds of correlated genetic variants in a single gene region. We demonstrate and explain why including too many genetic variants in such an analysis can lead to numerical instabilities and inflated Type 1 error rates. We also show that estimates based on a few variants can be highly sensitive to the choice of these variants. A novel approach is presented using principal components analysis to ensure that all variants contribute to the analysis, but without introducing numerical instabilities. We discuss practical implications of these findings for applied Mendelian randomization investigations. Software code in the R programming language for implementing the analyses discussed in the paper is provided in the Appendix.


\section*{Motivating example: serum testosterone and coronary heart disease risk}
We consider an example Mendelian randomization analysis with serum testosterone as the risk factor and coronary artery disease (CAD) risk as the outcome using genetic variants in the \emph{SHBG} gene region. The associations of 325 individual SNPs with testosterone are reported by \citet{jin2012}; associations of 322 of these variants with coronary artery disease risk are reported by the \citet{nikpey2015}. Previously, in an independent dataset, \citet{coviello2012} demonstrated at least six separate signals in the \emph{SHBG} gene region at a genome-wide level of significance, plus three more variants associated with sex hormone-binding globulin (SHBG) on adjustment for these six variants. In all analyses, correlations between variants are estimated using 1000 Genomes Phase 3 data on 503 individuals of European descent as reference data. A further 2 variants were monomorphic in the reference data; analyses are conducted using the remaining 320 variants. As variants in the \emph{SHBG} gene region are associated with circulating levels of both testosterone and SHBG, a positive Mendelian randomization finding would not distinguish which of these is a causal risk factor, but would suggest that sex hormone-related mechanisms have a causal role in cardiovascular disease.

Three approaches are taken here to choose which variants to include in a Mendelian randomization analysis. First, we take eight variants from the conditional analysis in the independent dataset reported by Coviello et al.\ (the association with testosterone in the data under analysis was not available for one variant). Second, we perform a stepwise conditional approach using the summarized associations reported by Jin et al., selecting at each step of the analysis the variant having the lowest p-value for association with the risk factor in the conditional analysis. We proceed until no additional variants are associated with the risk factor at $p < 0.0001$ or $p < 0.001$. This approach is implemented using the GCTA software. Third, we perform a stepwise pruning approach \citep{yang2012}, selecting at each step of the analysis the variant having the lowest p-value for association with the risk factor in a marginal (univariable) analysis. Once a variant is selected, all other variants whose correlation with the selected variant is greater in magnitude than a given correlation threshold (taken as 0.2, 0.4, 0.6, 0.8, 0.9, and 0.95; equivalent to an $r^2$ threshold of 0.04, 0.16, 0.36, 0.64, 0.81 and 0.9025) are removed from the analysis. We continue until each variant is either selected or removed. This ensures that a set of variants is chosen for each threshold correlation such that the variants are each marginally associated with the risk factor, and the pairwise correlations are all below the threshold correlation. Although a data-driven approach to selecting variants to include in a Mendelian randomization investigation is often unwise \citep{burgess2010avoiding}, in this case the associations with the risk factor and with the outcome are estimated in non-overlapping samples, and so ``winner's curse'' bias in the genetic associations with the outcome should not arise.

\begin{figure}[htb]
\centering
\parbox{0.48\textwidth}{ \includegraphics[height=7.5cm, bb = 0 0 504 504]{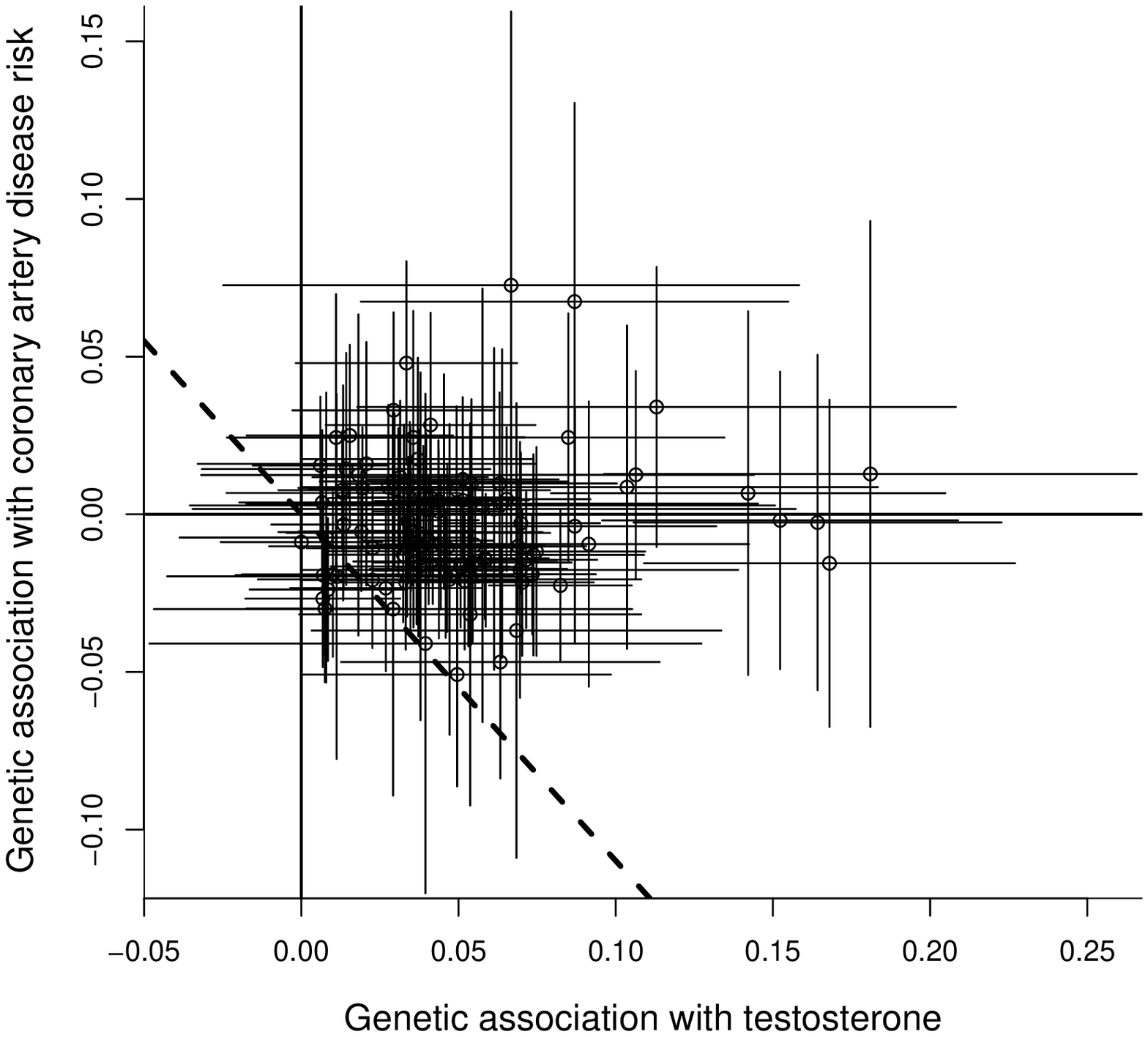} }%
\quad
\begin{minipage}{0.48\textwidth}%
\includegraphics[height=7.5cm, bb = 0 0 504 504]{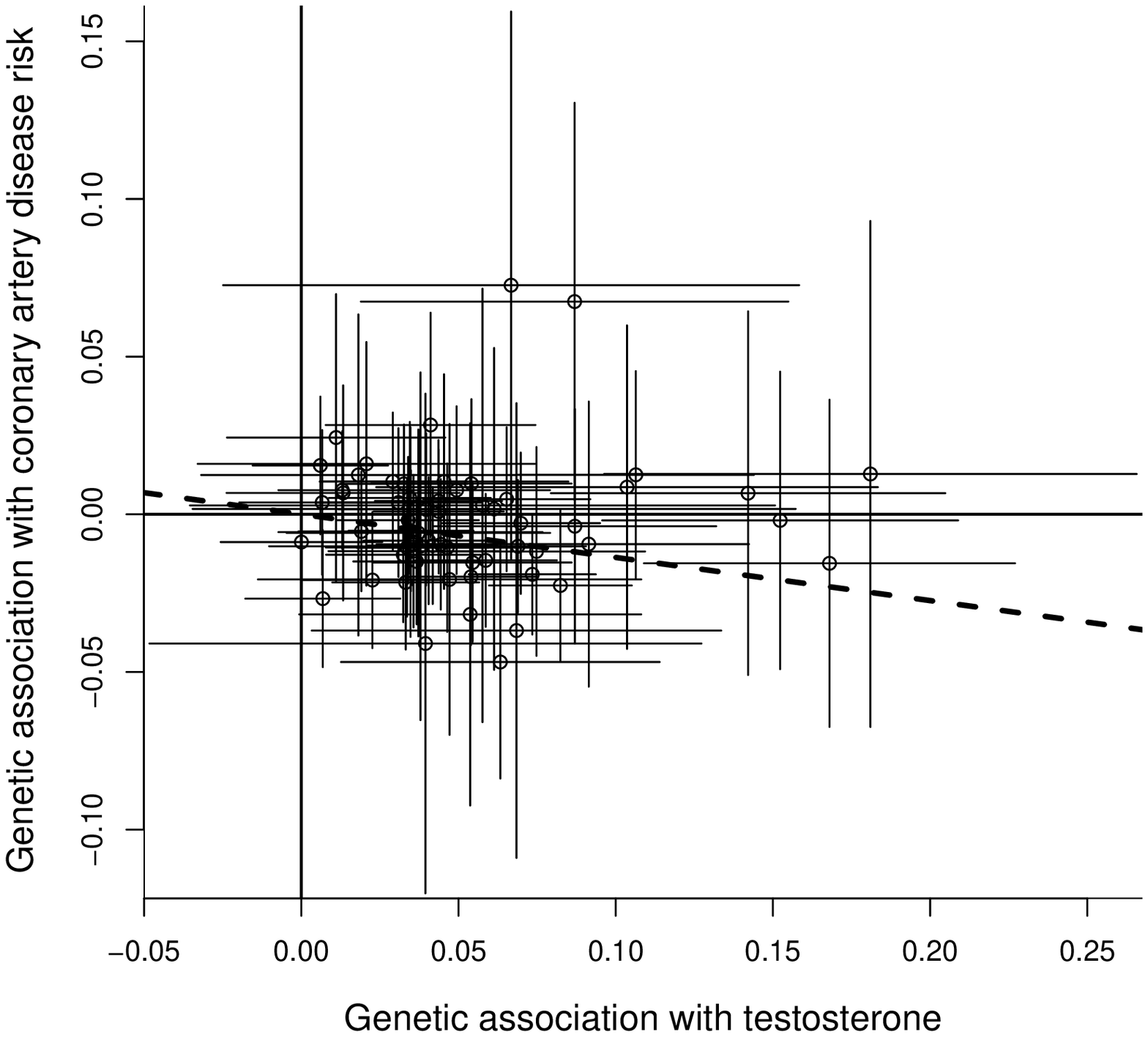}
\end{minipage}%
\caption{\onehalfspacing Estimated genetic associations and 95\% confidence intervals with testosterone (nmol/L, then log-transformed) and with coronary artery disease risk (log odds ratios): (left) for 104 genetic variants included in Mendelian randomization analysis with threshold correlation 0.95 ($r^2 = 0.9025$); (right) for 62 genetic variants with threshold correlation 0.8 ($r^2 = 0.64$). The heavy dashed line is the inverse-variance weighted estimate (accounting for correlation between variants).} \label{exres3}
\end{figure}

The Mendelian randomization estimates are presented in Table~\ref{exres2}. Fixed-effect analysis models that account for correlations between variants are used throughout. Despite the two approaches using a conditional analysis and the pruning approach at a threshold correlation of 0.2 including similar numbers of variants in the analysis, the causal estimates in these three analyses differed substantially -- by over two standard errors, and gave opposing substantive conclusions. In the pruning approach, as the threshold correlation increased, more variants were included in the Mendelian randomization analysis, and the precision of the causal estimate increased. However, for very large values of the threshold correlation, the standard error of the causal estimate is implausibly small. With a threshold correlation of 0.9, the standard error of the causal estimate was not defined due to the variance estimate being negative. With a threshold correlation of 0.95, the causal estimate is clearly spurious, as can be seen by visual inspection of the data (Figure~\ref{exres3}, left panel). The result with a correlation of 0.8 is also suspect (Figure~\ref{exres3}, right panel), as the variants having the greatest associations with testosterone all lie above the causal effect estimate. Even at lower threshold correlations of 0.4 and 0.6, the standard errors of the causal estimate are substantially lower than those calculated using the conditional approach. This may be due to the extra variants explaining additional variability in the risk factor; the reduction in standard error corresponds to a 97\% relative increase in variance explained by the variants at a threshold of 0.4 compared with at 0.2, and a 240\% increase at a threshold of 0.6. It is unclear which of the estimates in Table~\ref{exres2} are reliable, and therefore whether evidence supports testosterone as a causal risk factor for coronary heart disease risk or not.

\begin{table}[hbtp]
\begin{minipage}{\textwidth}
\begin{adjustwidth}{-0.4cm}{-0.4cm}
\setlength{\extrarowheight}{2pt}
\begin{center}
\begin{small}
\caption{\label{exres2} \textbf{Estimates in motivating example}}
\centering
\begin{tabular}[c]{ccccc}
\hline
Selection                      & \multicolumn{2}{c}{Threshold correlation}  & Number of &                \\
                               \cline{2-3}
approach                       & $\rho$ & $r^2$                             & variants  & Estimate (SE)  \\
\hline
Conditional analysis in        & \multirow{2}{*}{-} & \multirow{2}{*}{-}    & \multirow{2}{*}{8}
                                                                                        & \multirow{2}{*}{-0.258 (0.097)} \\
independent dataset (Coviello) &        &                                   &           &                \\
GCTA at $p < 0.0001$           &  -     &  -                                &    6      & -0.009 (0.058) \\
GCTA at $p < 0.001$            &  -     &  -                                &   19      & -0.068 (0.042) \\
Pruning                        & 0.2    & 0.04                              &    8      & -0.110 (0.094) \\
Pruning                        & 0.4    & 0.16                              &   20      & -0.085 (0.067) \\
Pruning                        & 0.6    & 0.36                              &   39      & -0.017 (0.051) \\
Pruning                        & 0.8    & 0.64                              &   62      & -0.137 (0.031) \\
Pruning                        & 0.9    & 0.81                              &   85      & -0.537 (-) \footnote{The variance estimate was negative, indicating that the weighting matrix was not positive definite, meaning that either the standard errors in the weighting matrix were imprecisely estimated, or else were not compatible with the correlation matrix.} \hphantom{.0}
                                                                                                       \\
Pruning                       & 0.95   & 0.9025                            &  104      & -1.099 (0.001) \\
\hline
\end{tabular}
\caption*{\onehalfspacing Estimates (standard errors, SE) of causal effect of testosterone on coronary artery disease risk (estimates are log odds ratios per unit increase in log-transformed testosterone) from inverse-variance weighted method (accounting for correlation) with variants selected using three different approaches and (for the pruning method) six different threshold correlations (measured by $\rho$ and by $r^2$).}
\end{small} 
\end{center}
\end{adjustwidth}
\end{minipage}
\end{table}

\section*{Choosing the right number of variants}
To resolve the question of how to choose which variants to include in a Mendelian randomization analysis, we explore reasons why analyses that include too many or too few genetic variants may go wrong, and propose a solution that incorporates associations on large numbers of genetic variants into the analysis, but does not suffer from numerical instabilities.

\subsection*{Too many variants: near-singular genetic correlation matrix}
A matrix is singular if it cannot be inverted -- formally, if the determinant of the matrix is zero. This occurs when the rows or columns of a matrix are linearly dependent; that is, at least one column (or row) can be calculated as a linear sum of multiples of the other columns (known as multicollinearity). This will occur for the genetic correlation matrix when two genetic variants are in perfect linkage disequilibrium, or alternatively if a small number of haplotypes are present in the data (perfect multicollinearity can occur even if no pair of variants is highly correlated). In contrast, a near-singular matrix can be inverted, but its determinant is close to zero. This occurs in a regression model when there is substantial, but not perfect, multicollinearity. As sample sizes for estimating genetic correlations increase, singular matrices will become less common, but near-singular genetic correlation matrices are likely to become more common. This is because a discrepant allele count in a single individual (which could represent a genotyping error) can lead to a singular matrix becoming non-singular. A near-singular matrix is problematic as elements of its inverse can be very large. In the motivating example with correlation thresholds of 0.9 and 0.95, the maximal element of the inverse of the correlation matrix is over 10 million.

If a matrix is exactly singular, then it cannot be inverted, and the analysis will report an error. If a matrix is near-singular, then the analysis may report an estimate without giving any indication that the estimate may be misleading (as observed in Figure~\ref{exres3}). In conjunction with discrepancies in the genetic association estimates, near-singular behaviour can lead to overly-precise as well as highly misleading estimates. Discrepancies may occur due to include rounding of association estimates (particularly for summarized genetic associations taken from the literature), inaccuracy and uncertainty in correlation estimates, and genetic association estimates and/or correlation estimates being estimated in different samples. When multiplied by the large numbers in the inverse of a near-singular genetic correlation matrix, small discrepancies in association estimates are magnified. Overprecision in the causal estimate will occur when genetic association estimates that should be similar based on the correlation matrix are more dissimilar than expected.

\subsection*{Too few variants: unstable estimates}
While theoretical considerations suggest that a Mendelian randomization analysis should be based on only variants associated with the risk factor in a conditional analysis, in practice this results in a Mendelian randomization estimate that only uses data on a small number of variants. In the motivating example, the conditional analyses suggest that less than 10 variants should be included in the analysis; associations with the remaining over 300 variants are ignored. In some cases and in particular in the motivating example, the causal estimate is highly sensitive to the choice of which variants are included in the analysis. This leads to unstable Mendelian randomization estimates -- if one of the selected variants in the conditional analysis happened not to be measured, or failed quality control (QC) criteria, then a different set of variants would have been obtained from the conditional analysis, resulting in a different Mendelian randomization estimate.

\subsection*{Just right?: principal components analysis}
One potential solution for resolving the problem of multiple correlated variants is principal components analysis (PCA). The use of PCA has been previously suggested for reducing the dimensionality of the instrumental variable space to resolve issues of weak instrument bias \citep{winkelried2011}, and as a tool for grouping variants in a fine-mapped gene region \citep{cai2013}. We perform unscaled PCA on a weighted version of the genetic correlation matrix  $\Psi_{j_1 j_2} = \hat{\beta}_{Xj1} \hat{\beta}_{Xj2}  \se(\hat{\beta}_{Yj1})^{-1} \se(\hat{\beta}_{Yj2})^{-1} \rho_{j_1 j_2}$. The diagonal elements of this matrix are the inverse-variance weights, and so each is equal to the precision of the causal estimate based on that variant alone.

Assuming that associations for all variants are estimated in the same sample size, these diagonal elements are proportional to the amount of variance in the risk factor explained by the genetic variant. This can be seen as the standard errors of the associations with the outcome will be directly proportional to the standard errors of the associations with the risk factor, which in turn relate to the minor allele frequencies $MAF_j$: if the variant is a diallelic SNP, then $\se(\hat{\beta}_{Xj})^{-2} \propto MAF_j (1-MAF_j)$ \citep{burgess2015scoretj}. (The proportion of variance in the risk factor explained by genetic variant $j$ is $\beta_{Xj}^2 \times MAF_j (1-MAF_j)$, where $\beta_{Xj}$ is measured in standard deviation units.) Hence, if the variants were uncorrelated, then the first principal component would be the genetic variant that explained the largest proportion of variance in the risk factor, and so on. For correlated variants, the first principal component represents a linear combination of variants that explains the largest proportion of variance in the risk factor, and each subsequent principal component is the linear combination of variants that explains the next largest proportion of variance while being orthogonal to the previous principal components.

This choice of matrix should be advantageous for Mendelian randomization investigations over PCA approaches on the unweighted matrix of genetic correlations. If two variants are perfectly correlated, but the estimates for one are measured in a larger sample size, then the precision of the association with the outcome ($\se(\hat{\beta}_{Yj})^{-1}$) will be greater for this variant, and so it will (correctly) be preferentially selected. The number of principal components to be included in the analysis can be chosen based on a threshold of variance in the weighted genetic correlation matrix. Once the principal components have been selected, we multiply the vector of genetic associations with the risk factor by the matrix of principal components, we multiply the vector of genetic associations with the outcome by the matrix of principal components, and pre- and post-multiply the genetic correlation matrix by the matrix of principal components. The IVW method is then performed on the transformed vectors of genetic associations and the transformed correlation matrix. If the matrix $\Psi = W \Lambda W^T$, where $W$ is the matrix of eigenvectors (or loadings), and $\Lambda$ is the diagonal matrix with the eigenvalues $\lambda_1 > \ldots > \lambda_J$ on the diagonal, then let $W_k$ be the matrix constructed of the first $k$ columns of $W$. Then we define:
\begin{align*}
\tilde{\boldsymbol\beta}_X &= W_k^T \hat{\boldsymbol\beta}_{X} \mbox{ as the transformed genetic associations with the risk factor} \\
\tilde{\boldsymbol\beta}_Y &= W_k^T \hat{\boldsymbol\beta}_{Y} \mbox{ as the transformed genetic associations with the outcome} \\
\tilde{\Omega} &= W_k^T \Omega W_k \mbox{ as the transformed weighting matrix.}
\end{align*}
Then the PCA-IVW estimate is given by:
\begin{equation}
(\tilde{\boldsymbol\beta}_{X}^T \tilde{\Omega}^{-1} \tilde{\boldsymbol\beta}_{X})^{-1} \tilde{\boldsymbol\beta}_{X}^T \tilde{\Omega}^{-1} \tilde{\boldsymbol\beta}_{Y}
\end{equation}

For the example of testosterone and CAD risk, 99\% of the variance in this matrix was explained by the first 8 principal components, and 99.9\% by the first 17 principal components. The corresponding estimates using these principal components as instruments were -0.065 (standard error 0.099) and -0.045 (0.083) respectively. These estimates are similar in precision to that using the previous conditional analysis for variable selection, but less precise than those calculated using the GCTA method on the data under analysis or a liberal correlation threshold in the pruning method.

\section*{Simulation study}
We illustrate statistical issues arising from using too many and too few variants in a series of simulation studies based on the motivating example. Again, fixed-effect analysis models are used throughout.

\subsection*{Sensitivity to choice of genetic variants}
First, we repeated the analyses of the motivating example except using only 180 of the 360 genetic variants at a time. This represents a scenario in which only a subset of the genetic variants in the analysis were measured. Sets of 180 variants were chosen at random 10\thinspace000 times.

\subsection*{Sensitivity to correlation matrix}
Second, we repeated the analyses of the motivating example except varying the correlation matrix. We took a bootstrap sample of the reference data (same size sample as the original data, sampled with replacement), and calculated a correlation matrix based on this sample. This procedure was performed 10\thinspace000 times.

For each of these simulation analyses, we performed the pruning method for selecting genetic variants at a threshold correlation of 0.2, 0.4, 0.6 and 0.8, and the PCA method using components that explained 99\% and 99.9\% of the variance in the summarized association matrix. Results are presented in Table~\ref{sim1}. In both simulation studies, as the threshold in the pruning approaches increased, the mean standard error of the causal estimates decreased, and the mean causal estimate also changed substantially. For a threshold correlation of $\rho = 0.8$, causal estimates were unstable, and were particularly sensitive to changes in the correlation matrix. In contrast, estimates using the PCA approach were not so precise, but they were far less variable between iterations.

\begin{table}[hbtp]
\begin{minipage}{\textwidth}
\begin{center}
\begin{footnotesize}
\setlength{\extrarowheight}{2pt}
\caption{\label{sim1} \textbf{Simulations varying choice of variants and correlation matrix}}
\centering
\begin{tabular}[c]{cccc@{\hskip 1em}ccc}
\hline
                          & \multicolumn{3}{c}{Varying choice of variants}  & \multicolumn{3}{c}{Varying correlation matrix} \\
                          \cmidrule(lr){2-4} \cmidrule(lr){5-7}
Selection approach        & Mean estimate &  SD     & Mean SE               & Mean estimate &  SD     & Mean SE              \\
\hline
Pruning at $\rho = 0.2$   & -0.100        &  0.044  &  0.094                & -0.114        &  0.035  &  0.090               \\
Pruning at $\rho = 0.4$   & -0.093        &  0.032  &  0.078                & -0.074        &  0.027  &  0.065               \\
Pruning at $\rho = 0.6$   & -0.009        &  0.049  &  0.060                & -0.018        &  0.052  &  0.046               \\
Pruning at $\rho = 0.8$   & -0.024        &  0.402  &  0.048 \footnote{Excluding 536 iterations in which the standard error was not defined.}
                                                                            &  - \footnote{Estimates were highly variable and the standard error was not defined for a large proportion of iterations.}
                                                                                            &  -      &       -              \\
PCA at 99\% of variance   & -0.053        &  0.028  &  0.098                & -0.051        &  0.027  &  0.096               \\
PCA at 99.9\% of variance & -0.045        &  0.025  &  0.084                & -0.047        &  0.017  &  0.083               \\
\hline
\end{tabular}
\caption*{\onehalfspacing Means of estimates, standard deviations (SD) of estimates, and mean standard errors (SE) for 10\thinspace000 iterations based on motivating example: i) varying the choice of variants and ii) varying the correlation matrix. Six approaches for selecting genetic variants are performed: four based on pruning at different correlation thresholds ($\rho$) and two based on principal components analysis (PCA).}
\end{footnotesize} 
\end{center}
\end{minipage}
\end{table}

\subsection*{Rounding of association estimates}
Finally, we simulated genetic associations with the risk factor and with the outcome directly. Genetic associations with the risk factor were drawn for 320 variants from a multivariable normal distribution with mean vector the measured genetic associations with testosterone from the motivating example and variance-covariance matrix $\Omega_X$, where $\Omega_{X j_1 j_2} = \se(\hat{\beta}_{Xj1}) \se(\hat{\beta}_{Xj2}) \rho_{j_1 j_2}$. The associations with the outcome are drawn from a multivariate normal distribution with mean zero and variance-covariance matrix $\Omega$, where $\Omega_{j_1 j_2} = \se(\hat{\beta}_{Yj1}) \se(\hat{\beta}_{Yj2}) \rho_{j_1 j_2}$ as defined above. This represents a null causal effect. We also set the mean of the distribution of the associations with the outcome as 0.1 times the associations with the risk factor, representing a causal effect of 0.1. We simulated 10\thinspace000 datasets for each value of the causal effect, and calculated the Mendelian randomization estimate using the same six approaches for variant selection as above. Additionally, we repeated the analyses but first rounding the genetic associations (and their standard errors) to three and two decimal places. 

\subsection*{Results}
Results are presented in Table~\ref{sim2} for the standard deviation of estimates, the mean standard error, and the empirical power of the 95\% confidence interval (the proportion of datasets in which the confidence interval excluded the null; this is the Type 1 error rate for a null causal effect). The mean estimates (not presented) were close to the true causal effect throughout for all approaches. As in the previous simulations, estimates from the pruning approach became more precise as the threshold correlation increased, although Type 1 error rates were above nominal levels for $\rho = 0.8$ even when the association estimates were not rounded. Rounding exacerbated false positive findings, and inflated Type 1 error rates were present in all methods when associations were rounded to 2 decimal places. Coverage rates were least affected when pruning at a threshold correlation of $\rho = 0.2$ or $0.4$ and for the PCA approaches. With a positive causal effect, power increased as the threshold increased, although judging estimators by power estimates is misleading when Type 1 error rates are inflated. Power of the PCA approaches was similar to that using a pruning threshold of $\rho = 0.2$, and was greater using principal components that explained a greater proportion of the weighted correlation matrix.

\setlength{\tabcolsep}{4pt}
\begin{table}[hbtp]
\begin{minipage}{\textwidth}
\begin{adjustwidth}{-0.8cm}{-0.8cm}
\begin{center}
\begin{footnotesize}
\setlength{\extrarowheight}{2pt}
\caption{\label{sim2} \textbf{Simulation rounding association estimates}}
\centering
\begin{tabular}[c]{cccccccccc}
\hline
                          & \multicolumn{3}{c}{Unrounded}  & \multicolumn{3}{c}{3 decimal places}  & \multicolumn{3}{c}{2 decimal places} \\
                          \cmidrule(lr){2-4} \cmidrule(lr){5-7} \cmidrule(lr){8-10}
Selection approach        &  SD   & Mean SE & Power        &  SD  & Mean SE  & Power               &  SD   & Mean SE & Power              \\
\hline
\multicolumn{10}{c}{Null causal effect}                                                                                                   \\
\hline
Pruning at $\rho = 0.2$   & 0.080 & 0.079   &  5.0         & 0.080 & 0.080   &  4.9                & 0.086 & 0.077   &  7.3               \\
Pruning at $\rho = 0.4$   & 0.067 & 0.066   &  5.0         & 0.067 & 0.066   &  5.1                & 0.073 & 0.063   &  9.2               \\
Pruning at $\rho = 0.6$   & 0.049 & 0.049   &  5.0         & 0.050 & 0.050   &  4.9                & 0.066 & 0.047   & 16.5               \\
Pruning at $\rho = 0.8$   & 0.027 & 0.022   & 10.5         & 0.175 & 0.022   & 40.8                & 0.418 & 0.020   & 62.2               \\
PCA at 99\% of variance   & 0.089 & 0.090   &  4.6         & 0.090 & 0.090   &  4.6                & 0.094 & 0.083   &  8.0               \\
PCA at 99.9\% of variance & 0.075 & 0.075   &  4.6         & 0.075 & 0.076   &  4.5                & 0.079 & 0.069   &  9.0               \\
\hline
\hline
\multicolumn{10}{c}{Positive causal effect of 0.1}                                                                                        \\
\hline
Pruning at $\rho = 0.2$   & 0.080 & 0.079   & 24.8         & 0.080 & 0.080   & 24.6                & 0.086 & 0.077   & 27.9               \\
Pruning at $\rho = 0.4$   & 0.067 & 0.066   & 33.6         & 0.067 & 0.066   & 33.2                & 0.073 & 0.063   & 37.0               \\
Pruning at $\rho = 0.6$   & 0.049 & 0.049   & 54.3         & 0.050 & 0.050   & 51.9                & 0.066 & 0.047   & 53.1               \\
Pruning at $\rho = 0.8$   & 0.027 & 0.022   & 88.8         & 0.172 & 0.022   & 86.7                & 0.644 & 0.020   & 79.3               \\
PCA at 99\% of variance   & 0.089 & 0.090   & 19.6         & 0.090 & 0.090   & 19.5                & 0.095 & 0.083   & 25.1               \\
PCA at 99.9\% of variance & 0.075 & 0.075   & 26.1         & 0.075 & 0.076   & 25.6                & 0.079 & 0.069   & 32.6               \\
\hline
\end{tabular}
\caption*{\onehalfspacing Standard deviations (SD) of estimates, mean standard errors (SE), and empirical power based on the 95\% confidence interval for 10\thinspace000 simulated datasets using six approaches for selecting genetic variants. Results are also given on rounding the association estimates to a fixed number of decimal places.} 
\end{footnotesize}
\end{center}
\end{adjustwidth}
\end{minipage}
\end{table}
\setlength{\tabcolsep}{6pt}

\section*{Discussion}
As the cost of high-density genome sequencing continues to fall, additional signals are likely to be identified within known loci. There will be growing demand for methods to exploit correlated instruments in Mendelian randomization, as the addition of correlated variants can improve power to detect a causal effect. In this paper, we first connected previously known results together to show from theoretical arguments that genetic variants included in a Mendelian randomization analysis should be those that are associated with the risk factor in a conditional analysis. If the variants are combined in an allele score, then the conditional (multivariable) associations with the risk factor should be used as weights in the allele score to obtain the most efficient analysis. If only summarized data are available, then the same analysis can be replicated with the marginal (univariable) associations using an extension to the inverse-variance weighted method to account for correlations between variants.

However, difficulties arise when there are many correlated genetic variants in a single gene region that are associated with the risk factor (fine-mapping genetic data). Including too few genetic variants in an analysis means that estimates are less precise, but also highly variable, in that different approaches to choosing variants can lead to markedly different estimates. However, including too many variants can lead to numerical instabilities and overly precise estimates with inflated Type 1 error rates. These numerical instabilities are not computational issues, but arise due to inconsistencies in the data: for example, if association estimates are rounded to a fixed number of decimal places, or if association or correlation estimates are obtained in different samples. It is difficult in practice to judge at what threshold these numerical issues begin to occur, although in the simulation examples considered, problems regularly occurred when pruning variants at a threshold correlation of 0.8 ($r^2 = 0.64$), and occasionally occurred at a threshold correlation of 0.6 ($r^2 = 0.36$). We note as well that $r^2$ is not always a good measure of correlation between genetic variants; near-singular matrices can occur when the pairwise correlations measured by $r^2$ are low but there are haplotypes represented in the data, or when the minor allele frequencies of variants differ but a common variant `tags' a rare variant (high D-prime, but low $r^2$).

As an alternative approach, we have proposed a method for selecting instruments based on principal components analysis of a weighted version of the genetic correlation matrix. This approach constructs instruments as linear combinations of genetic variants. As the linear combinations are orthogonal, the approach does not suffer as much with respect to numerical instabilities. Additionally, the method incorporates data on all the genetic variants into the analysis, and consequently causal estimates from the approach are less variable. Estimates from the principal components analysis approach are less precise than those from the variable selection approaches considered here (GCTA and pruning); however, they are less variable with respect to choices of how to implement the analysis (in particular the choice of variants).

\subsection*{Comparison with previous work}
The inverse-variance weighted method presented here is a simple application of generalized weighted linear regression, and is not unique to Mendelian randomization. The same method has been used in a variety of contexts including discovery genetics \citep{zhu2016}, and prediction and model selection \citep{chen2015, benner2016, newcombe2016}. A number of different solutions have been proposed to the problem of highly-correlated variants, including pruning and clumping at a threshold correlation, and adding a small positive number to the diagonal of the correlation matrix \citep{gusev2016}. In the applied example of the paper at a correlation threshold of $\rho = 0.8$, adding 0.1 to the diagonal of the correlation matrix changed the causal estimate from -0.137 (standard error 0.031) to -0.065 (0.057). Although the substantial change in the causal estimate is indicative of near-singular behaviour, it would seem preferable for estimation to simply use a stricter correlation threshold rather than misspecifying the correlation matrix (and better still to use the principal component approach presented in this manuscript).

We believe that Mendelian randomization differs somewhat from other analysis contexts, as an instrumental variable analysis relies on inferences from a single gene region (for example, for a protein risk factor where the gene region is the coding region for the risk factor) or a small number of gene regions. Another feature of Mendelian randomization is the prevalence of the summarized data and two-sample settings, in which discrepancies in genetic associations are likely to arise.

Principal components approaches have been suggested before for fine-mapping data, with Wallace demonstrating that 70\% of the variance in the genetic correlation matrix could be explained by an average of 7 components for 49 test gene regions \citep{wallace2013}. A key innovation here is weighting the genetic correlation matrix, meaning that principal components with the greatest eigenvalues will be those that explain the most variance in the risk factor. This means that it is more likely that an analysis based on a small number of principal components will have reasonable power to detect a causal effect. For example, if there is only one causal variant in the gene region, then 100\% of the variance would be explained by one principal component, even if there were other uncorrelated variants in the gene region.

We advocate the principal components analysis method proposed in this paper as a worthwhile approach to analyse fine-mapped genetic data for Mendelian randomization.

\vspace{3mm}

\subsection*{Acknowledgements}
We would like to thank Paul Newcombe and Chris Wallace (both MRC Biostatistics Unit, Cambridge) for comments on an earlier version of this manuscript. Stephen Burgess is supported by Sir Henry Dale Fellowship jointly funded by the Wellcome Trust and the Royal Society (Grant Number 204623/Z/16/Z).


\bibliographystyle{apalike} 
\bibliography{masterref}

\begin{thebibliography}{}

\bibitem[Benner et~al., 2016]{benner2016}
Benner, C., Spencer, C.~C., Havulinna, A.~S., Salomaa, V., Ripatti, S., and
  Pirinen, M. 2016.
\newblock {FINEMAP: Efficient variable selection using summary data from
  genome-wide association studies}.
\newblock {\em Bioinformatics}, 32(10):1493--1501.

\bibitem[Borenstein et~al., 2009]{borenstein2009}
Borenstein, M., Hedges, L., Higgins, J., and Rothstein, H. 2009.
\newblock {\em {Introduction to meta-analysis. Chapter 34: Generality of the
  basic inverse-variance method}}.
\newblock Wiley.

\bibitem[Burgess et~al., 2013]{burgess2013genepi}
Burgess, S., Butterworth, A.~S., and Thompson, S.~G. 2013.
\newblock {Mendelian randomization analysis with multiple genetic variants
  using summarized data}.
\newblock {\em Genetic Epidemiology}, 37(7):658--665.

\bibitem[Burgess et~al., 2015a]{burgess2014pleioajeappendix}
Burgess, S., Dudbridge, F., and Thompson, S.~G. 2015a.
\newblock {Re: ``Multivariable Mendelian randomization: the use of pleiotropic
  genetic variants to estimate causal effects''}.
\newblock {\em American Journal of Epidemiology}, 181(4):290--291.

\bibitem[Burgess et~al., 2016]{burgess2015scoretj}
Burgess, S., Dudbridge, F., and Thompson, S.~G. 2016.
\newblock {Combining information on multiple instrumental variables in
  Mendelian randomization: comparison of allele score and summarized data
  methods}.
\newblock {\em Statistics in Medicine}, 35(11):1880--1906.

\bibitem[Burgess et~al., 2015b]{burgess2014twosample}
Burgess, S., Scott, R., Timpson, N., Davey~Smith, G., Thompson, S.~G., and
  {EPIC-InterAct Consortium} 2015b.
\newblock {Using published data in Mendelian randomization: a blueprint for
  efficient identification of causal risk factors}.
\newblock {\em European Journal of Epidemiology}, 30(7):543--552.

\bibitem[Burgess and Thompson, 2015]{burgess2015book}
Burgess, S. and Thompson, S.~G. 2015.
\newblock {\em {Mendelian randomization: methods for using genetic variants in
  causal estimation}}.
\newblock Chapman \& Hall.

\bibitem[Burgess et~al., 2011]{burgess2010avoiding}
Burgess, S., Thompson, S.~G., and {CRP CHD Genetics Collaboration} 2011.
\newblock {Avoiding bias from weak instruments in Mendelian randomization
  studies}.
\newblock {\em International Journal of Epidemiology}, 40(3):755--764.

\bibitem[Cai et~al., 2013]{cai2013}
Cai, M., Dai, H., Qiu, Y., et~al. 2013.
\newblock {{SNP} set association analysis for genome-wide association studies}.
\newblock {\em PLOS One}, 8(5):e62495.

\bibitem[{CARDIoGRAMplusC4D Consortium}, 2015]{nikpey2015}
{CARDIoGRAMplusC4D Consortium} 2015.
\newblock {A comprehensive 1000 Genomes-based genome-wide association
  meta-analysis of coronary artery disease}.
\newblock {\em Nature Genetics}, 47:1121--1130.

\bibitem[Chen et~al., 2015]{chen2015}
Chen, W., Larrabee, B.~R., Ovsyannikova, I.~G., et~al. 2015.
\newblock {Fine mapping causal variants with an approximate Bayesian method
  using marginal test statistics}.
\newblock {\em Genetics}, 200(3):719--736.

\bibitem[Coviello et~al., 2012]{coviello2012}
Coviello, A.~D., Haring, R., Wellons, M., et~al. 2012.
\newblock {A genome-wide association meta-analysis of circulating sex
  hormone--binding globulin reveals multiple Loci implicated in sex steroid
  hormone regulation}.
\newblock {\em PLoS Genetics}, 8(7):e1002805.

\bibitem[Davey~Smith and Ebrahim, 2003]{daveysmith2003}
Davey~Smith, G. and Ebrahim, S. 2003.
\newblock {`Mendelian randomization': can genetic epidemiology contribute to
  understanding environmental determinants of disease?}
\newblock {\em International Journal of Epidemiology}, 32(1):1--22.

\bibitem[Didelez and Sheehan, 2007]{didelez2007}
Didelez, V. and Sheehan, N. 2007.
\newblock {Mendelian randomization as an instrumental variable approach to
  causal inference}.
\newblock {\em Statistical Methods in Medical Research}, 16(4):309--330.

\bibitem[Gusev et~al., 2016]{gusev2016}
Gusev, A., Ko, A., Shi, H., et~al. 2016.
\newblock {Integrative approaches for large-scale transcriptome-wide
  association studies}.
\newblock {\em Nature Genetics}, 48(3):245--252.

\bibitem[Inoue and Solon, 2010]{inoue2010}
Inoue, A. and Solon, G. 2010.
\newblock {Two-sample instrumental variables estimators}.
\newblock {\em The Review of Economics and Statistics}, 92(3):557--561.

\bibitem[Jin et~al., 2012]{jin2012}
Jin, G., Sun, J., Kim, S.-T., et~al. 2012.
\newblock {Genome-wide association study identifies a new locus JMJD1C at 10q21
  that may influence serum androgen levels in men}.
\newblock {\em Human Molecular Genetics}, 21(23):5222--5228.

\bibitem[Johnson, 2013]{johnson2013vig}
Johnson, T. 2013.
\newblock {Efficient calculation for multi-SNP genetic risk scores}.
\newblock Technical report, The Comprehensive R Archive Network.
\newblock Available at
  http://cran.r-project.org/web/packages/gtx/vignettes/ashg2012.pdf [last
  accessed 2017/4/19].

\bibitem[Lawlor et~al., 2008]{lawlor2007}
Lawlor, D., Harbord, R., Sterne, J., Timpson, N., and Davey~Smith, G. 2008.
\newblock {Mendelian randomization: using genes as instruments for making
  causal inferences in epidemiology}.
\newblock {\em Statistics in Medicine}, 27(8):1133--1163.

\bibitem[Newcombe et~al., 2016]{newcombe2016}
Newcombe, P.~J., Conti, D.~V., and Richardson, S. 2016.
\newblock {JAM: A scalable Bayesian framework for joint analysis of marginal
  SNP effects}.
\newblock {\em Genetic Epidemiology}, 40(3):188--201.

\bibitem[Pierce et~al., 2011]{pierce2010}
Pierce, B., Ahsan, H., and VanderWeele, T. 2011.
\newblock {Power and instrument strength requirements for Mendelian
  randomization studies using multiple genetic variants}.
\newblock {\em International Journal of Epidemiology}, 40(3):740--752.

\bibitem[Wallace, 2013]{wallace2013}
Wallace, C. 2013.
\newblock {Statistical testing of shared genetic control for potentially
  related traits}.
\newblock {\em Genetic Epidemiology}, 37(8):802--813.

\bibitem[Winkelried and Smith, 2011]{winkelried2011}
Winkelried, D. and Smith, R.~J. 2011.
\newblock {Principal components instrumental variable estimation}.
\newblock Technical report, Faculty of Economics, University of Cambridge.
  Cambridge Working Papers in Economics 1119.

\bibitem[Wooldridge, 2009]{wooldridge2009ch15}
Wooldridge, J. 2009.
\newblock {\em {Introductory econometrics: A modern approach. Chapter 15:
  Instrumental variables estimation and two stage least squares}}.
\newblock South-Western, Nashville, TN.

\bibitem[Yang et~al., 2012]{yang2012}
Yang, J., Ferreira, T., Morris, A., et~al. 2012.
\newblock {Conditional and joint multiple-SNP analysis of GWAS summary
  statistics identifies additional variants influencing complex traits}.
\newblock {\em Nature Genetics}, 44(4):369--375.

\bibitem[Zhu et~al., 2016]{zhu2016}
Zhu, Z., Zhang, F., Hu, H., et~al. 2016.
\newblock {Integration of summary data from GWAS and eQTL studies predicts
  complex trait gene targets}.
\newblock {\em Nature Genetics}, 48:481--487.

\end{thebibliography}

\clearpage

\section*{Figure legends}
\indent \indent \textbf{Figure \ref{exres3}:} Estimated genetic associations and 95\% confidence intervals with testosterone (nmol/L, then log-transformed) and with coronary artery disease risk (log odds ratios): (left) for 104 genetic variants included in Mendelian randomization analysis with threshold correlation 0.95 ($r^2 = 0.9025$); (right) for 62 genetic variants with threshold correlation 0.8 ($r^2 = 0.64$). The heavy dashed line is the inverse-variance weighted estimate (accounting for correlation between variants).

\clearpage
\renewcommand{\thesection}{A\arabic{section}}
\renewcommand{\thesubsection}{A.\arabic{subsection}}
\renewcommand{\thetable}{A\arabic{table}}
\renewcommand{\thefigure}{A\arabic{figure}}
\setcounter{table}{0}
\setcounter{figure}{0}
\renewcommand{\tablename}{Web Table}
\renewcommand{\figurename}{Web Figure}
\setcounter{section}{0}
\setcounter{subsection}{0}
\section*{Appendix}
\subsection{Software code}
We provide R code to implement the methods discussed in this paper. The genetic variants are represented by \texttt{g} (a matrix of allele counts for the genetic variants), the risk factor by \texttt{x}, and the outcome by \texttt{y}. Weights for the allele score are represented by \texttt{wts}. The associations of the candidate instruments with the risk factor are denoted \texttt{betaXG} with standard errors \texttt{sebetaXG}. The associations of the candidate instruments with the outcome are denoted \texttt{betaYG} with standard errors \texttt{sebetaYG}. With a continuous outcome, these associations are usually estimated using linear regression; with a binary outcome, using logistic regression.

\normalsize{The two-stage least squares (2SLS) method can be implemented using the \emph{sem} package:}\par
\scriptsize{
\begin{verbatim}
library(sem)
beta_2sls = tsls(y, cbind(x, rep(1,parts)), cbind(g, rep(1,parts)),
                  w=rep(1, parts))$coef[1]
                     # w are the weights in the two-stage least squares method
                     #  (w is set to one for all individuals)
                     # the cbind(..., rep(1,parts)) ensures that a constant term is
                     #  included in both regression stages of the 2SLS method
se_2sls   = sqrt(tsls(y, cbind(x, rep(1,parts)), cbind(g, rep(1,parts)),
                  w=rep(1, parts))$V[1,1])
\end{verbatim}
}\par

\normalsize{
Genetic variants can be collapsed into an allele score, and the score can be used in the 2SLS method: }\par
\scriptsize{
\begin{verbatim}
library(sem)
score = g%*%wts
beta_score = tsls(y, cbind(x, rep(1,parts)), cbind(score, rep(1,parts)),
                  w=rep(1, parts))$coef[1]
se_score = sqrt(tsls(y, cbind(x, rep(1,parts)), cbind(score, rep(1,parts)),
                  w=rep(1, parts))$V[1,1])
\end{verbatim}
}\par
\normalsize{
If the genetic variants are perfectly uncorrelated, and the weights are the coefficients from univariable regression analyses of the risk factor on each of the genetic variants in turn, then these two analyses are equivalent.}\par

\normalsize{
Inverse-variance weighted estimate (ignoring correlation): }\par
\scriptsize{
\begin{verbatim}
beta_IVW      = summary(lm(betaYG~betaXG-1, weights=sebetaYG^-2))$coef[1]
se_IVW.fixed  = summary(lm(betaYG~betaXG-1, weights=sebetaYG^-2))$coef[1,2]/
                 summary(lm(betaYG~betaXG-1, weights=sebetaYG^-2))$sigma
se_IVW.random = summary(lm(betaYG~betaXG-1, weights=sebetaYG^-2))$coef[1,2]/
             max(summary(lm(betaYG~betaXG-1, weights=sebetaYG^-2))$sigma,1)
\end{verbatim}
}\par
\normalsize{
Although fixed-effects models are used throughout this paper, the (multiplicative) random-effects analysis is preferred when heterogeneity between the causal estimates from each genetic variant is expected (provided that there are enough genetic variants in the model to obtain a reasonable estimate of the heterogeneity). Heterogeneity would generally be expected when using genetic variants from multiple gene regions that may have different mechanisms of influencing the risk factor, but not when using multiple variants in the same gene region that should have similar mechanisms of effect.}\par

\normalsize{
Inverse-variance weighted estimate (accounting for correlation): }\par
\scriptsize{
\begin{verbatim}
Omega               = sebetaYG%o%sebetaYG*rho
beta_IVWcorrel      = solve(t(betaXG)%*%solve(Omega)%*%betaXG)*t(betaXG)%*%solve(Omega)%*%betaYG
se_IVWcorrel.fixed  = sqrt(solve(t(betaXG)%*%solve(Omega)%*%betaXG))
resid               = betaYG-beta_IVWcorrel*betaXG
se_IVWcorrel.random = sqrt(solve(t(betaXG)%*%solve(Omega)%*%betaXG))*
                       max(sqrt(t(resid)%*%solve(Omega)%*%resid/(length(betaXG)-1)),1)
\end{verbatim}
}\par
\normalsize{
The matrix \texttt{rho} comprises the pairwise correlations between the genetic associations (in particular, the genetic associations with the outcome). Provided that are genetic associations estimated in the same participants, these are equal to the correlations between the genetic variants themselves.}\par

\vspace{3mm}

\normalsize{
Inverse-variance weighted estimate (accounting for correlation) using principal components: }\par
\scriptsize{
\begin{verbatim}
Phi     = (betaXG/sebetaYG)%o%(betaXG/sebetaYG)*rho
summary(prcomp(Phi, scale=FALSE))
K       = which(cumsum(prcomp(Phi, scale=FALSE)$sdev^2/sum((prcomp(Phi, scale=FALSE)$sdev^2)))>0.99)[1]
       # K is number of principal components to include in analysis
       # this code includes principal components to explain 99% of variance in the risk factor
betaXG0 = as.numeric(betaXG%*%prcomp(Phi, scale=FALSE)$rotation[,1:K])
betaYG0 = as.numeric(betaYG%*%prcomp(Phi, scale=FALSE)$rotation[,1:K])
Omega   = sebetaYG%o%sebetaYG*rho
pcOmega = t(prcomp(Phi, scale=FALSE)$rotation[,1:K])%*%Omega%*%prcomp(Phi, scale=FALSE)$rotation[,1:K]

beta_IVWcorrel.pc     = solve(t(betaXG0)%*%solve(pcOmega)%*%betaXG0)*t(betaXG0)%*%solve(pcOmega)%*%betaYG0
se_IVWcorrel.fixed.pc = sqrt(solve(t(betaXG0)%*%solve(pcOmega)%*%betaXG0))
\end{verbatim}
}\par

\normalsize{
The inverse-variance weighted method accounting for correlation can also be performed using the standard linear regression command after weighting the data by a Cholesky decomposition: }\par
\scriptsize{
\begin{verbatim}
Omega    = sebetaYG%o%sebetaYG*rho
c_betaXG = solve(t(chol(Omega)))%*%betaXG
c_betaYG = solve(t(chol(Omega)))%*%betaYG

beta_IVWcorrel      = lm(c_betaYG~c_betaXG-1)$coef[1]
se_IVWcorrel.fixed  = sqrt(1/(t(betaXG)%*%solve(Omega)%*%betaXG))
se_IVWcorrel.random = sqrt(1/(t(betaXG)%*%solve(Omega)%*%betaXG))*max(summary(lm(c_betaYG~c_betaXG-1))$sigma,1)
\end{verbatim}
}\par

\normalsize{\hphantom{end}}
\clearpage

\subsection{Proof of equality of 2SLS and inverse-variance weighted estimates}
\textbf{Variants uncorrelated:} If the we write the risk factor as $X$ (usually an $N \times 1$ matrix, although the result can be generalized for multiple risk factors), the outcome as $Y$ (an $N \times 1$ matrix), and the instrumental variables as $Z$ (an $N \times J$ matrix), then the two-stage least squares estimate of causal effects is:
\begin{equation}
\hat{\beta}_{2SLS} = [X^T Z (Z^T Z)^{-1} Z^T X]^{-1} X^T Z (Z^T Z)^{-1} Z^T Y. \notag
\end{equation}
This estimate can be obtained by sequential regression of the risk factor on the instrumental variables, and then the outcome on fitted values of the risk factor from the first-stage regression.

Regression of $Y$ on $Z$ gives beta-coefficients $\hat{\beta}_Y = (Z^T Z)^{-1} Z^T Y$ with standard errors the square roots of the diagonal elements of the matrix $(Z^T Z)^{-1} \sigma^2$ where $\sigma$ is the residual standard error. If the instrumental variables are perfectly uncorrelated, then the off-diagonal elements of $(Z^T Z)^{-1} \sigma^2$ are all equal to zero. Regression of $X$ on $Z$ gives beta-coefficients $\hat{\beta}_X = (Z^T Z)^{-1} Z^T X$. Weighted linear regression of the beta-coefficients $\hat{\beta}_Y$ on the beta-coefficients $\hat{\beta}_X$ using the inverse-variance weights $(Z^T Z) \sigma^{-2}$ gives an estimate:
\begin{align}
        \hphantom{=}& [\hat{\beta}_X^T (Z^T Z) \hat{\beta}_X]^{-1} \sigma^{-2} \hat{\beta}_X^T (Z^T Z) \sigma^2 \hat{\beta}_Y \notag \\
                  =& [X^T Z (Z^T Z)^{-1} (Z^T Z) (Z^T Z)^{-1} Z^T X]^{-1} X^T Z (Z^T Z)^{-1} (Z^T Z) (Z^T Z)^{-1} Z^T Y \notag \\
                  =& [X^T Z (Z^T Z)^{-1} Z^T X]^{-1} X^T Z (Z^T Z)^{-1} Z^T Y \notag \\ 
                  =& \hat{\beta}_{2SLS} \notag
\end{align}
The assumption of uncorrelated instrumental variables ensures that the regression coefficients from univariate regressions (as in the regression-based methods) equal those from multivariable regression (as in the two-stage least squares method).

\textbf{Variants correlated:} If the variants are correlated, then the same argument holds, except that the weights in the weighted linear regression of the beta-coefficients $\hat{\beta}_Y$ on the beta-coefficients $\hat{\beta}_X$ are $(Z^T Z) P \sigma^{-2}$, where $P$ is the (symmetric) correlation matrix.

\end{document}